\documentclass[twocolumn,aps,prb,showpacs,floatfix,superscriptaddress,unsortedaddress]{revtex4}
\usepackage{epsfig}

\newcommand{\beq}{\begin{equation}}
\newcommand{\eeq}{\end{equation}}
\newcommand{\etal}{{\it et al.}}
\newcommand{\bea}{\begin{eqnarray}}
\newcommand{\eea}{\end{eqnarray}}
\begin{document}

\title{Raman scattering in a d-wave superconductor - a one gap scenario}
\author{A. V. Chubukov}
\affiliation{Department of Physics, University of Wisconsin, Madison, WI
53706}
\author{M. R. Norman}
\affiliation{
Materials Science Division, Argonne National Laboratory, Argonne,
Illinois 60439}
\date{\today}
\begin{abstract}
Recent Raman scattering data in $B_{1g}$ and $B_{2g}$ geometries in the superconducting state of underdoped cuprates were interpreted as evidence for two distinct energy gaps. 
 We argue that these data can be equally well explained within a one gap scenario if 
final state interactions are taken into account. In particular, we show 
 that they can account for the differing doping dependences
 of the Raman peaks in these two geometries.
\end{abstract}

\maketitle

\section{Introduction}

A key issue in the physics of cuprates is the relation between the pseudogap and the 
superconducting gap~\cite{AP}.  One class of theories implies that
 the pseudogap and the superconducting gap are of different origin. The pseudogap appears at $T^*$ in the antinodal region as a result of some competing instability (with or without long range order), with a magnitude that increases with underdoping, scaling with $T^*$. The superconducting gap, on the other hand, appears only below $T_c$ on an arc of the Fermi surface around the node that was not removed by the pseudogap, and has a magnitude that decreases with underdoping, scaling with $T_c$.  This can be contrasted with another class of theories that assumes there is a single
 $d-$wave pairing gap (not necessarily of the form $\cos k_xa - \cos k_ya$),  whose magnitude scales with $T^*$ 
both in the nodal and antinodal regions.  Above $T_c$, superconducting coherence is destroyed by fluctuations, but gap-like features in the spectral function
 survive as long as the magnitude of the fermionic damping, $\eta(T)$, is smaller than the
 angle-dependent pairing self-energy, $\Delta(\phi)$
(specifically, $|\Delta (\phi)| > \sqrt{3} \eta (T)$~\cite{mike_mod,elihu}).   This give rise to a $T$
dependent arc around the node, where along the arc,
 $|\Delta (\phi)|$ is smaller than $\sqrt{3} \eta (T)$, and
  the spectral function has a peak at the Fermi energy, as in the normal state.

One gap scenarios are consistent with recent photoemission measurements of the temperature dependence of the Fermi arc above $T_c$~\cite{amit1}.
Measurements by the same group below $T_c$~\cite{amit2} found a simple d-wave gap of
the form $\cos 2\phi$ with a magnitude
and temperature dependence unrelated to $T_c$, but consistent with $T^*$.
Similar behavior has been inferred from Fourier transforms of recent scanning tunneling
data~\cite{davis}, except for regions near the node where there is some evidence for a suppressed
gap as in earlier photoemission data~\cite{mesot}.  A 
 one gap scenario is also consistent with point contact SIS tunneling data~\cite{john}.

Recent Raman studies of underdoped $Hg-$based
 cuprates by Le Tacon \etal~\cite{gabi}  were interpreted as evidence for the existence of two distinct energy scales in the underdoped cuprates.  The Raman spectra in $B_{1g}$ and $B_{2g}$ geometries behave differently with doping: the peak in $B_{1g}$ geometry shifts to a higher energy with underdoping and, to a first approximation, tracks (twice) the antinodal gap, while the peak in 
$B_{2g}$ geometry shows the same doping dependence as $T_c$ and 
shifts to a lower energy with underdoping.  As $B_{1g}$ and $B_{2g}$ 
responses predominantly come from the antinodal and nodal regions, respectively~\cite{girsh}, this result  was interpreted as evidence
 that the nodal and antinodal gaps have different doping dependences. This does not necessary imply two distinct gaps (the interpretation of Ref.~\onlinecite{gabi} assumes a $d-$wave gap with doping
 dependent anisotropy as in Ref.~\onlinecite{mesot}), 
but still it does imply that the gap function behaves differently in these two momentum regions.

Here, we argue that the Raman data can be explained equally well within a one gap scenario, as a result of  final state interactions, which are assumed to increase with underdoping. We find that
final state interactions can lead to different doping dependences 
 of the $B_{1g}$ and $B_{2g}$ responses even if the superconducting gap has a
 simple $\cos 2\phi$ form.

\section{Formalism}

Without final state interactions,  the Raman intensity due to electronic excitations
 is given by the Fermi surface average of the sum of normal ($GG$) 
and anomalous ($FF$)  bubbles, weighted with Raman vertices~\cite{RMP}.  In a situation where the self-energy depends only on $\omega$, but not on $k$ normal to the Fermi surface (apart from a trivial velocity renormalization), the Raman intensity at $T=0$ 
in the geometry labeled by `i' is  the imaginary part of the Raman bubble: $R_i (\Omega)$, which for $\Omega >0$ is given by
\begin{widetext}
\beq
R_i(\Omega) =  R_0 \left< \gamma_i^2 (\phi) \left[2 - \int_{-\infty}^\infty d \omega   \frac{\sqrt{{\tilde\omega}^2_+ - \Delta^2(\phi)}\sqrt{{\tilde\omega}^2_- - \Delta^2(\phi)} - {\tilde\omega}_+ {\tilde\omega}_- + \Delta^2 (\phi)}{\sqrt{{\tilde\omega}^2_+ - \Delta^2(\phi)} \sqrt{{\tilde\omega}^2_- - \Delta^2(\phi)} (\sqrt{{\tilde\omega}^2_+ - \Delta^2(\phi)} + \sqrt{{\tilde\omega}^2_- -\Delta^2 (\phi)})}\right]\right>_{FS}       
\label{2}
\eeq
\end{widetext}    
where $R_0$ is a normalization factor,
${\tilde \omega}_{\pm} = \omega_{\pm} - \Sigma (\omega_{\pm})$, 
$\omega_{\pm} = \omega \pm \Omega/2$, $\phi$ is the angular variable along the Fermi surface, $<...>_{FS}$ denotes averaging over the Fermi surface
( $<...>_{FS} = (2/\pi)\int_0^{\pi/2} ... d \phi$ for  a circular Fermi surface),
and $\Delta (\phi)$ is the pairing self-energy which for simplicity we assume to be independent of frequency.
For a circular Fermi surface, we assume a $d-$wave gap of the form
 $\Delta (\phi) = \Delta \cos 2\phi$. The vertices $\gamma_i$ are different for different scattering geometries and are  $\gamma_{B_{1g}} (\phi) = \cos 2\phi$ (i.e., $\cos k_xa - \cos k_ya$), and $\gamma_{B_{2g}} (\phi) = \sin 2\phi$ (i.e., $\sin k_xa$ $\sin k_ya$). Because of the angular dependences of $\gamma_i$, $B_{1g}$ Raman scattering predominantly probes the antinodal regions, where $\cos 2\phi$ is the largest, while the $B_{2g}$ Raman intensity comes from the nodal regions, where $\sin 2\phi$ is the largest~\cite{girsh,RMP,devereaux}.  

 In the BCS approximation, $\Sigma =0$, and the Raman intensity reduces to~\cite{klein,devereaux} 
\beq
Im [R_i(\Omega)] = 4\pi R_0 Re \left[\left< \frac{\gamma^2_i (\phi) 
\Delta^2 (\phi)}{\Omega \sqrt{\Omega^2 - 4 \Delta^2 (\phi)}}\right>_{FS}\right]
\label{r_1}
\eeq
The expressions for $Im R_{B_{1g}}$ and $Im R_{B_{2g}}$ in this approximation can be analytically expressed in terms of complete elliptic integrals~\cite{devereaux}. At small frequencies, 
only nodal fermions contribute to the Raman intensity, and 
\beq
Im R_{B_{1g}} (\Omega) = \frac{3 \pi}{8} R_0 \left(\frac{\Omega}{v_n}\right)^3, ~~
Im R_{B_{2g}} (\Omega) = \frac{\pi}{2}~R_0 \frac{\Omega}{v_n}
\label{2_1}
\eeq
where $v_n$  is the nodal `velocity' ($v_n = 2\Delta$ for a $\cos 2\phi$ gap).
The real parts of the $R_{B_{1g}}$ and $R_{B_{2g}}$ bubbles behave as
\beq
Re R_{B_{1g}} (\Omega) = \frac{1}{3} R_0 
\left(\frac{\Omega}{\Delta}\right)^2,~~
Re R_{B_{2g}} (\Omega) = -\frac{1}{3} R_0 
\left(\frac{\Omega}{\Delta}\right)^2.
\label{2_11}
\eeq

At larger frequencies,
 $Im  R_{B_{1g}} (\Omega)$ 
 diverges logarithmically at $2\Delta$, while $Re  R_{B_{1g}} (\Omega)$ 
jumps from a positive to a negative value:
\bea
&&Im  R_{B_{1g}} (\Omega) = R_0 \log{\frac{2\Delta}{|2\Delta -\Omega|}}, \nonumber \\
&&Re  R_{B_{1g}} (2\Delta + 0) = -R_0 \left(\pi - \frac{4}{3}\right), \nonumber \\
&&Re  R_{B_{1g}} (2\Delta - 0) = \frac{4R_0}{3}.
\label{b1g}
\eea
The $B_{2g}$ intensity ($Im  R_{B_{2g}} (\Omega)$) 
has a broad maximum  at around $1.6 \Delta$. The real part of the $B_{2g}$ bubble is negative for all frequencies, and has a weak minimum at $2\Delta$, where  $Re  R_{B_{2g}} (\Omega) =-4R_0/3$. 
  
\begin{figure}[h]
\epsfig{file=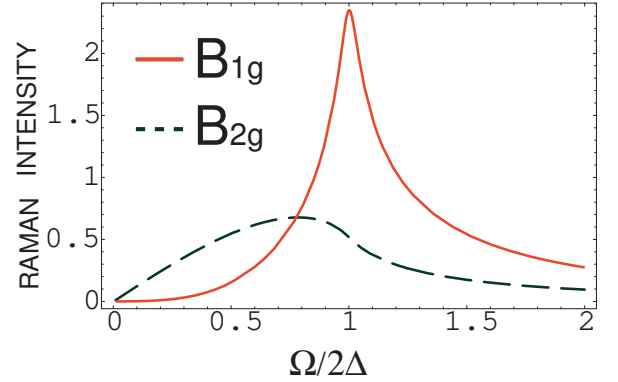,width=7.8cm}
\caption{(Color online)
The imaginary
 part of the  Raman bubble, $B_{1g}$ (solid line) and $B_{2g}$ (dashed line),
 with $\Omega$ in units of $2\Delta$.
 In the BCS approximation, these  are the $B_{1g}$ and $B_{2g}$ Raman intensities.
The logarithmic  divergence of the $B_{1g}$ intensity at $2\Delta$
is smoothed by the presence of a small damping, $\eta =0.06 \Delta$.
Here and in other figures, we plot the Raman intensity in units of $4 R_0/\pi$.
} \label{fig1}
\end{figure}

\begin{figure}[h]
\epsfig{file=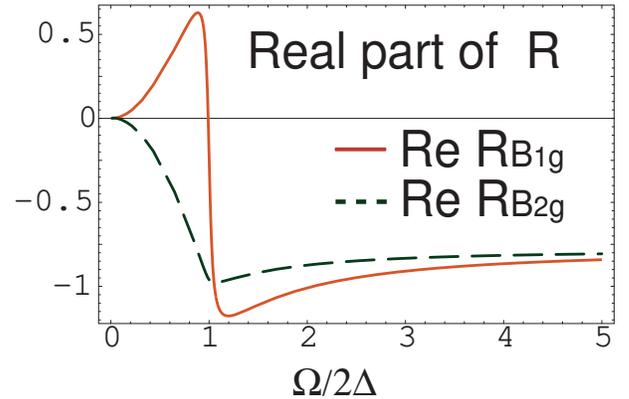,width=8.cm}
\caption{(Color online) 
The real part of the  Raman bubble, $B_{1g}$ (solid line) and $B_{2g}$ (dashed line). Observe that 
Re$R_{B_{1g}}$ changes sign, but Re$R_{B_{2g}}$ remains negative for all frequencies.  
} \label{fig:re}
\end{figure}

The imaginary and real parts of $R_{B_{1g}}$ and $R_{B_{2g}}$ for general $\Omega$ are shown in 
Figs.~\ref{fig1} and \ref{fig:re}.  

Final state interactions arise from multiple insertions of the fermion-fermion interaction into the Raman bubble. The fully renormalized
 four-fermion interaction $\Gamma_{\alpha\beta,\gamma\delta} (q, \Omega)$ 
generally has components  in the spin and charge channels, and depends on 
 the transferred momenta $q=k-k'$ and transferred frequency $\Omega = \omega - \omega'$.
  Restricting to only $B_{1g}$ and $B_{2g}$ harmonics, we can approximate $\Gamma$ as 
\begin{widetext}
\beq
\Gamma_{\alpha\beta,\gamma\delta} = \delta_{\alpha\beta} \delta_{\gamma \delta}
\left(\Gamma^{c}_{B_{1g}} \gamma_{B_{1g}} (k)  \gamma_{B_{1g}} (k')
+ \Gamma^{c}_{B_{2g}} \gamma_{B_{2g}} (k)  \gamma_{B_{2g}} (k') \right) 
+ \sigma_{\alpha\beta} \sigma_{\gamma \delta} 
\left(\Gamma^{s}_{B_{1g}} \gamma_{B_{1g}} (k)  \gamma_{B_{1g}} (k')
+ \Gamma^{s}_{B_{2g}} \gamma_{B_{2g}} (k)  \gamma_{B_{2g}} (k') \right) 
\label{6}
\eeq
\end{widetext}
where indices c and s refer to charge and spin, and $\Gamma^{c,s} = \Gamma^{c,s} (\omega-\omega')$. 

We make two assumptions in evaluating the effect of final state interactions. First, we assume that the random phase approximation (RPA) is valid, with the full Raman intensity given by
\begin{equation}
 Im\left[\frac{R_i (\Omega)}{1 + \Gamma_i R_i (\Omega)}\right] = 
\frac{Im R_i (\Omega)}{(1+ \Gamma_i Re R_i (\Omega))^2 + (\Gamma_i Im R_i (\Omega))^2}
\label{7}
\eeq
where $\Gamma_i = \Gamma^{c}_i + 3 \Gamma^{s}_i$. 
Second, we assume that the interactions $\Gamma^{c,s} (\Omega)$ depend only weakly on frequency for $\Omega << W$, where $W$ is the fermion bandwidth,
 but are strongly reduced at frequencies comparable to $W$. This holds if, e.g., the effective interaction is mediated by overdamped spin fluctuations.  
This is relevant to final state interactions because 
 the constant term in $R_i$ (the `2' in the r.h.s.~of Eq.~(\ref{2})) comes from fermions with energies comparable to $W$, while the rest comes from fermions with energies of order $\Omega << W$. Because the
 effective interaction is strongly reduced at $\Omega \sim W$, the
 constant term should be dropped from  $Re R_i$ in the
 the denominator of Eq.~(\ref{7}). At the same time, for
 the rest of $R_i$, the interactions $\Gamma^{i}$ can be safely approximated by constants \cite{foot1}.

The effect of the final state interactions obviously depends on the sign of $\Gamma_i$.  The interaction in the $B_{1g}$ channel is the same one as 
 gives rise to 
 $d-$wave superconductivity.  If the spin component of $\Gamma_{B_{1g}}$ 
dominates over the charge component, $\Gamma_{B_{1g}}$ is negative~\cite{cbm,cdk}. $\Gamma_{B_{2g}}$ on the other hand  is not related to pairing, and 
 in general can have either sign. 

\section{Results}

\subsection{General $d-$wave gap}

As discussed above, the peak positions in the $B_{1g}$ and $B_{2g}$
 Raman intensities as a function of doping do not scale with each other.  One way to account for this is to still use the BCS formula, but assume that the gap $\Delta (\phi)$ progressively deviates from the simple $\cos 2\phi$ form with underdoping, such that the gap near
 the node probed in $B_{2g}$ scattering, and the antinodal gap probed in $B_{1g}$ scattering, have different doping dependences (a `two scale' scenario). A
 simple  way to account for this is to add a $\cos 6 \phi$ harmonic to the $d_{x^2-y^2}$ gap 
 function~\cite{mesot,gabi}:
\beq
\Delta (\phi) = \Delta \left[(1-a) \cos2\phi + a \cos 6 \phi\right]
\label{10}
\eeq
Note that the maximum value of the gap is still $\Delta$.
In Figs.~\ref{fig11} and \ref{fig10}, we show  the $B_{1g}$ and $B_{2g}$ Raman 
profiles for $a=0$ and a finite $a$ versus $\Omega$.  We see that the
 peak position in $B_{2g}$ geometry progressively shifts to  lower frequencies
with increasing $a$, and flattens. The $B_{1g}$ peak, on the other hand, doesn't move, but just broadens.  Because $\Delta$ increases with underdoping, the $B_{1g}$ peak, located at $2\Delta$, actually shifts  to higher frequency with underdoping. The $B_{2g}$ peak, however, 
shifts to lower frequency with underdoping if the doping dependence of $a$ 
 overshadows the growth of $\Delta$. We also note that the slope of the $B_{2g}$ Raman intensity at $\Omega=0$ increases with increasing $a$.
\begin{figure}[h]
\epsfig{file=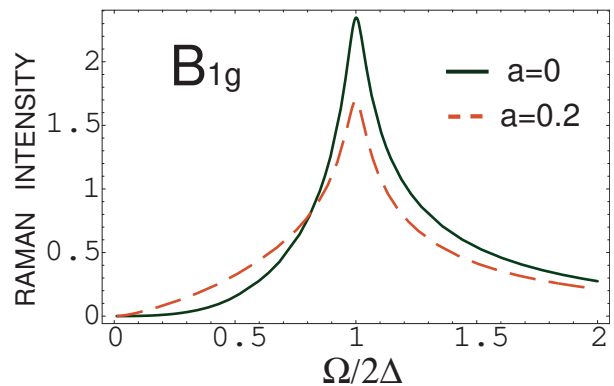,width=8.cm}
\caption{(Color online) 
$B_{1g}$ Raman intensity 
for a gap anisotropy parameter $a=0$ (solid) and $a=0.2$ (dashed), with
 $\eta = 0.06 \Delta$.  Observe that the peak doesn't move but just broadens with $a$.} \label{fig11}
\end{figure}
\begin{figure}[h]
\epsfig{file=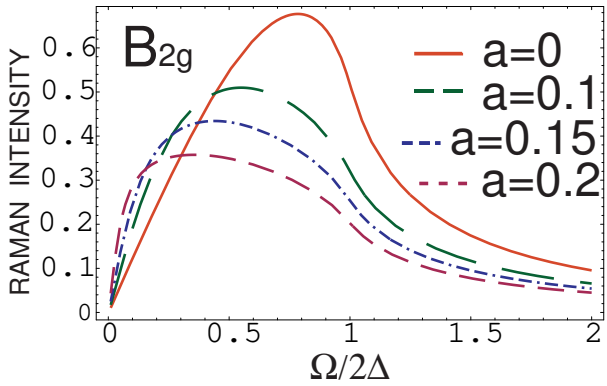,width=8.cm}
\caption{(Color online) 
 $B_{2g}$ Raman intensity for a gap anisotropy parameter $a=0, 0.1, 0.15, 0.2$, with $\eta = 0.06 \Delta$.
The peak moves to lower frequency with increasing $a$.} \label{fig10}
\end{figure}

\subsection{Final state interactions}
 
We now show that final state interactions also lead to distinct behaviors
 of the $B_{1g}$ and $B_{2g}$ Raman intensities (Figs.~5-8), 
even for a simple $d-$wave gap (i.e., $a=0$). 
In Fig.~\ref{fig4} we show the 
 $B_{1g}$ Raman intensity for negative $\Gamma_{B_{1g}}$.  
From here on, we 
express $\Gamma$ in units of $\pi/(4R_0)$.
  As $|\Gamma_{B_{1g}}|$ increases, 
 the $B_{1g}$ peak shifts to a somewhat lower frequency compared to 
$2\Delta$, and its intensity decreases.
 Still,  the actual frequency of the 
$B_{1g}$ peak will increase with underdoping if the increase of $\Delta$ overshadows the reduction of the peak frequency relative to $2\Delta$. 

For  $\Gamma_{B_{1g}} >0$, the effect of the final state interaction is opposite -- the peak shifts to a higher frequency, and its intensity becomes quite large for   $\Gamma_{B_{1g}} \sim 1$.
  For larger  $\Gamma_{B_{1g}}$, the intensiy drops. We show this behavior in Fig.~\ref{fig:n1}.  
\begin{figure}[h]
\epsfig{file=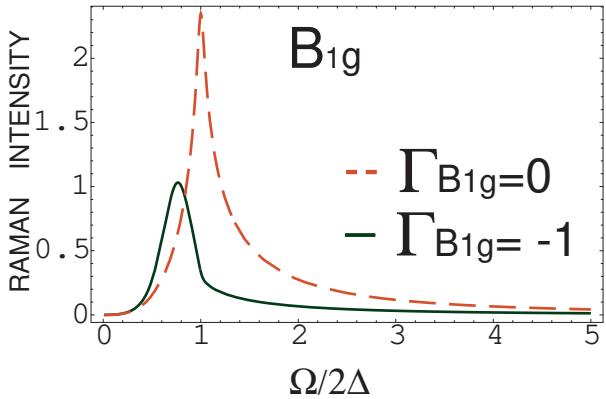,width=8.cm}
\caption{(Color online) 
$B_{1g}$ Raman intensity for
  $\Delta (\phi) = \Delta \cos 2\phi$ with final state interaction
$\Gamma_{B_{1g}} =0$ (dashed) and 
$\Gamma_{B_{1g}} =-1$ (solid).
For finite  $\Gamma_{B_{1g}}$, the intensity decreases in magnitude, and the peak shifts to a somewhat lower frequency compared to $2\Delta$.
Here, and in other figures, $\Gamma$ is expressed in units of $\pi/4R_0$.} \label{fig4}
\end{figure}
\begin{figure}[h]
\epsfig{file=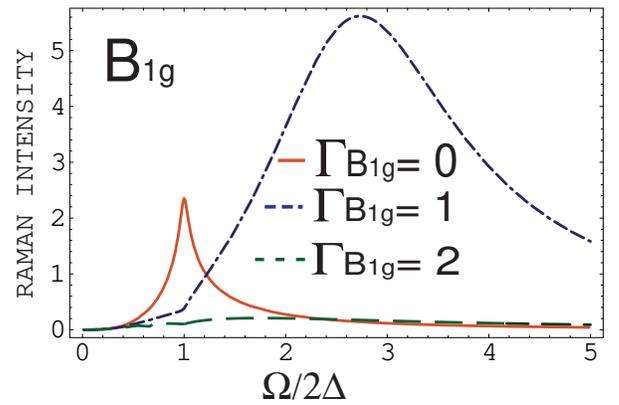,width=8.cm}
\caption{(Color online) 
$B_{1g}$ Raman intensity with the final state interaction of opposite sign compared to Fig.~\protect\ref{fig4}. 
 The peak shifts to a higher frequency and its
 intensity increases
for $\Gamma_{B_{1g}} \sim 1$. For
 larger $\Gamma_{B_{1g}}$, the intensity drops.} \label{fig:n1}
\end{figure}
\begin{figure}[h]
\epsfig{file=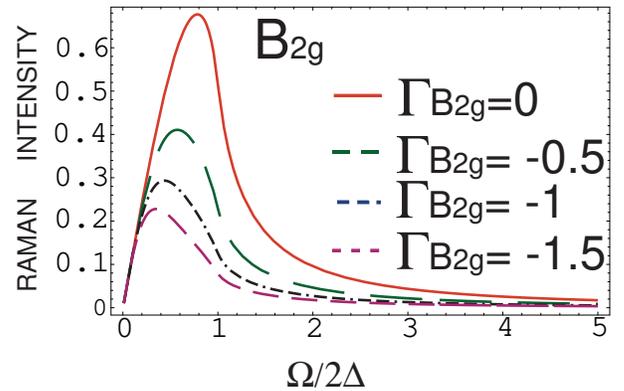,width=8.cm}
\caption{(Color online) 
$B_{2g}$ Raman intensity with the final state interaction
$\Gamma_{B_{2g}} =0, -0.5, -1, -1.5$.  With increasing $\Gamma_{B_{2g}}$,
the peak progressively shifts down in $\Omega$ and the intensity decreases. The
 slope at $\Omega=0$, however, remains unchanged.} \label{fig4_1}
\end{figure}
\begin{figure}[h]
\epsfig{file=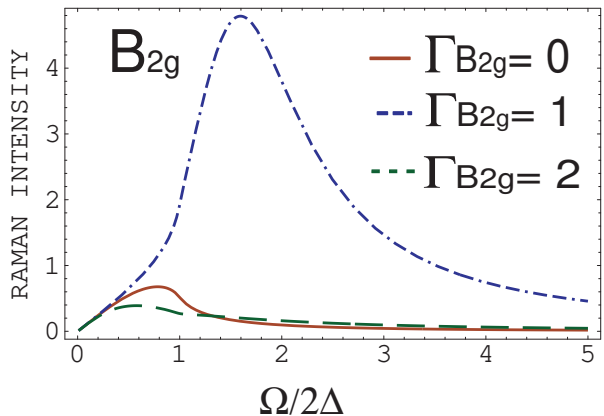,width=8.cm}
\caption{(Color online) 
$B_{2g}$ Raman intensity with the final state interaction of opposite sign compared to Fig.~\protect\ref{fig4_1}.
 The peak shifts to a higher frequency and its
 intensity increases for $\Gamma_{B_{2g}} \sim 1$ 
(the same effect as in Fig. \protect\ref{fig:n1}). 
 For larger $\Gamma_{B_{2g}}$, the intensity drops. } \label{fig4_11}
\end{figure}

We found a stronger effect of the final state interaction in $B_{2g}$ geometry. For negative  $\Gamma_{B_{2g}}$,  the intensity decreases, and the 
 peak frequency rapidly shifts to a smaller value, demonstrating the same trend as with the gap anisotropy parameter, $a$. We show this behavior in Fig.~\ref{fig4_1}. 
 The only qualitative difference between this figure and
 Fig.~\ref{fig10} is that in Fig.~\ref{fig4_1},
 the slope of the $B_{2g}$ Raman intensity 
is unaffected by the final state interaction simply 
because near $\Omega =0$, `1' dominates over $R(\Omega)$ in the denominator of Eq.~(\ref{7}).  

For positive  $\Gamma_{B_{2g}}$ the effect is the opposite -- the peak shifts
 to a larger frequency, and its intensity increases quite dramatically for 
 $\Gamma_{B_{2g}} \sim 1$ (see Fig.~\ref{fig4_11}).
  For larger  $\Gamma_{B_{2g}}$, the intensity drops, and the profile becomes almost flat.  

\subsubsection{Theoretical reasoning}

The behavior of the $B_{1g}$ intensity for a $d-$wave gap in the presence 
 of final state interactions has been studied before~\cite{cbm,cdk,dahm}.
The reduction of the peak frequency to below $2\Delta$ for negative $\Gamma_{B_{1g}}$ was understood as an exciton-like effect, similar to the one which gives rise to a peak in the dynamic spin susceptibility in a $d_{x^2-y^2}$ superconductor. $Re R_{B_{1g}} (\Omega)$ is positive below $2\Delta$ and is a monotonic function of  $\Omega$ in this range (Fig.~\ref{fig:re}).  The denominator of Eq.~(\ref{7}) contains 
$1 + \Gamma_{B_{1g}} Re R_{B_{1g}} (\Omega)$, and for strong enough, negative 
$\Gamma_{B_{1g}}$ (from Eq.~(\ref{b1g}),  $|\Gamma_{B_{1g}}| > 3/\pi$),
$1 + \Gamma_{B_{1g}} Re R_{B_{1g}} (\Omega)$ equals zero 
somewhere below $2\Delta$. In the absence of damping, this would imply
 a pole, the same as for the spin susceptibility.
For the Raman intensity, however, $Im R_{B_{1g}} (\Omega)$ is non-zero for all frequencies. Still, 
$Im  R_{B_{1g}} (\Omega)$ scales as $\Omega^3$ and is generally small below $2\Delta$.
The net result is that the
`2$\Delta$' peak shifts  to a lower frequency with increasing $|\Gamma_{B_{1g}}|$~\cite{cbm,cdk}.
For large  enough $|\Gamma_{B_{1g}}|$, the peak position and the peak intensity scale as
$\sqrt{-1/\Gamma}$, as can be easily derived from Eqs.~(\ref{2_1}), (\ref{2_11}), and (\ref{7}).
We note that the intensity of the peak
 decreases compared to the case with no final state interactions.
For  $|\Gamma_{B_{1g}}| < 3/\pi$,
a pseudo-resonance does not occur. Still, we find the
peak position and peak intensity monotonically decrease with
increasing $|\Gamma_{B_{1g}}|$.

For the opposite sign of $\Gamma_{B_{1g}}$ ($\Gamma_{B_{1g}} >0$),  there also exists a pseudo-resonance, but 
 this time for $\Omega >2\Delta$ (Fig.~\ref{fig:n1}). The reason is that $Re R_{B_{1g}} (\Omega)$ changes sign and becomes negative above $2\Delta$
 (Fig.~\ref{fig:re}). For positive $\Gamma_{B_{1g}} \sim 1$, $1 + 
\Gamma_{B_{1g}} Re R_{B_{1g}} (\Omega)$ crosses zero at some 
$\Omega$ above $2\Delta$, for which $Im R_{B_{1g}} (\Omega)$ is again small (Fig.~\ref{fig1}).  
For larger positive  $\Gamma_{B_{1g}}$, a pseudo-resonance does not develop, and the intensity drops because of the presence of $\Gamma^2_{B_{1g}}$ in the denominator of  Eq.~(\ref{7}).

For the $B_{2g}$ case,
 $Re R_{B_{2g}} (\Omega)$ at small frequencies has the opposite sign
(Eq.~(\ref{2_11})).  Applying the same logic as above,
 one would then expect a shift of the $B_{2g}$  peak to 
a lower frequency for $\Gamma_{B_{2g}}>0$, as in this case
 $1 + \Gamma_{B_{2g}} Re R_{B_{2g}} (\Omega)$ crosses zero below $2\Delta$.
However, Figs.~\ref{fig4_1} and \ref{fig4_11} show
 the opposite trend - the peak shifts to a higher frequency for $\Gamma_{B_{2g}} \sim 1$, and to a lower frequency for $\Gamma_{B_{2g}} <0$.
 The explanation is that at small frequencies,
 the imaginary part of $R_{B_{2g}} (\Omega)$ scales linearly with $\Omega$, and  is much larger than
  $Im R_{B_{1g}} (\Omega) \propto \Omega^3$.  Because 
 $Im R_{B_{2g}} (\Omega)$ is large, Eq.~(\ref{7}) is not 
 enhanced even when $ \Gamma_{B_{2g}} Re R_{B_{2g}} (\Omega) =-1$.
In this situation, the trend with 
 positive $\Gamma_{B_{2g}}$ is determined 
 by the fact that $ ReR_{B_{2g}} (\Omega)$ doesn't change sign with $\Omega$ -- it passes through a weak minimum at $\Omega =2\Delta$ and remains negative at higher frequencies. For $\Gamma_{B_{2g}} \sim 1$,  the $B_{2g}$ intensity has a pseudo-resonance at $\Omega >2\Delta$, just like the $B_{1g}$ intensity does. As the imaginary part of the Raman bubble 
$Im R_{B_{2g}} (\Omega)$ is small above $2\Delta$ (Fig.~\ref{fig1}), the intensity of this pseudo-resonance is large, and it dominates the profile of the Raman intensity 
for $\Gamma_{B_{2g}} \sim 1$. For larger $\Gamma_{B_{2g}}$, a pseudo-resonance does not develop ( $|\Gamma_{B_{2g}} Re R_{B_{2g}} (\Omega)| >1$), and the intensity drops, similar to the $B_{1g}$ intensity.

For negative $\Gamma_{B_{2g}}$, a pseudo-resonance does not occur, and the trend with  $\Gamma_{B_{2g}} <0$ is determined by the fact that
 the final state interaction monotonically  decreases the intensity of the  $B_{2g}$ Raman response as the frequency increases. At the smallest $\Omega$, this effect is vanishingly small, but for $\Omega \sim 2\Delta$, the intensity is reduced quite substantially. As a result, the $B_{2g}$ Raman intensity develops a maximum at a frequency which becomes smaller as $|\Gamma_{B_{2g}}|$ increases, 
and the intensity at the maximum progressively decreases, as shown in Fig.~\ref{fig4_1}.

\subsection{Fermionic self-energy}

For completeness, we also analyzed the effect of the fermionic self-energy 
  on the Raman profile. As in previous work~\cite{mike_mod,elihu},
 we assumed that the self-energy $\Sigma (\omega)$ can be approximated by 
 $\Sigma (\omega)  = -i \eta \rm{sgn}(\omega)$. In the supercoducting state, $\eta$ is small, but it rapidly increases above $T_c$ and can 
 affect the Raman profile in the pseudogap regime.  

 The results for the $\cos 2\phi$ gap are shown in Figs.~\ref{fig8} and \ref{fig7}.  The peaks in both geometries get broader when damping increases. In addition,  the $B_{1g}$ intensity increases at small $\Omega$  with increasing $\eta$, 
while the $B_{2g}$ intensity decreases at small $\Omega$.  
We verified that the
 same trend presists for a general d-wave gap (i.e., $a$ non-zero), and when 
 final state interactions are included. 
\begin{figure}[h]
\epsfig{file=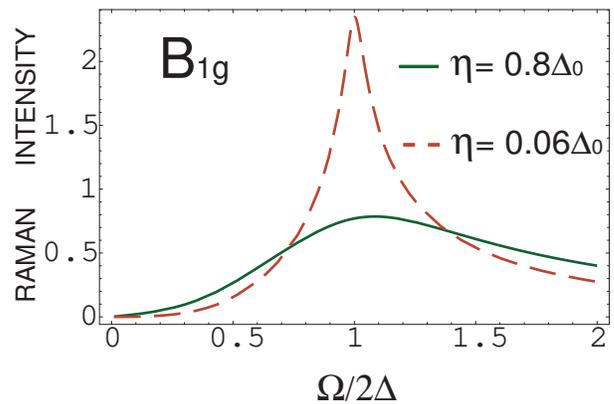,width=8.cm}
\caption{(Color online) 
$B_{1g}$ Raman intensity  for $\eta = 0.06 \Delta$ (dashed) 
and $\eta = 0.8 \Delta$ (solid).
With increasing $\eta$, the peak broadens, and the intensity at small frequencies increases.} \label{fig8}
\end{figure}
\begin{figure}[h]
\epsfig{file=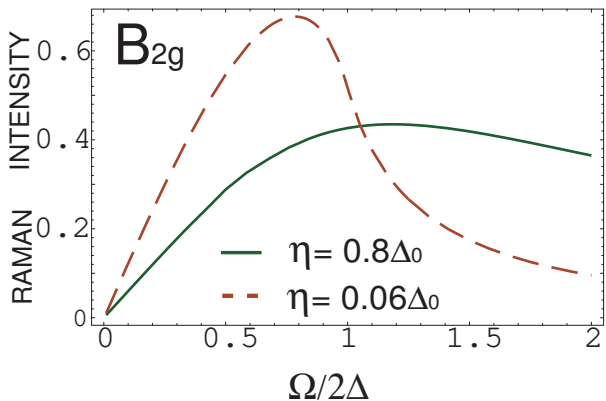,width=8.cm}
\caption{(Color online) 
 $B_{2g}$ Raman intensity for $\eta = 0.06 \Delta$ (dashed) 
and $\eta = 0.8 \Delta$ (solid).
 The peak broadens and the slope at $\Omega =0$ decreases.} \label{fig7}
\end{figure}

\subsubsection{Universal slope for $B_{2g}$ intensity}

We note  in passing that a finite $\eta$ gives rise to a new
 universal regime in the $B_{2g}$ response. Namely, for small $\Omega < \eta << \Delta$, the $B_{2g}$ Raman intensity is linear in $\Omega$ and depends only on the nodal velocity, but is different from that in Eq.~(\ref{2_1}).   
\beq
Im R_{B_{2g}} (\Omega) = \frac{4}{\pi} R_0 \frac{\Omega}{v_n}, ~~\Omega < \eta
\label{3}
\eeq
This universality of $Im R_{B_{2g}} (\Omega)$ has the  same origin as the 
universal conductivity of a $d-$wave superconductor~\cite{lee}.  
We found, however, that this universal behavior sets in only for small $\eta < 0.1 \Delta$, while at larger $\eta$, the  $B_{2g}$ Raman profile at small
 frequencies is almost independent of the nodal velocity.

\section{Comparison with experiment}

 We demonstrated that there are two possible ways to explain the Raman data.
 One explanation assumes that the gap anisotropy changes with underdoping. 
 Another explanation is that the distinct behavior of the $B_{1g}$ and $B_{2g}$ 
 peaks is the effect of final state interactions 
 (Figs.~\ref{fig4} and ~\ref{fig4_1}).
 In both cases, one can reproduce the doping dependences of the $B_{1g}$ and $B_{2g}$ peaks. The second explanation is consistent with recent photoemission data that  finds a simple d-wave
gap~\cite{amit2}, even in underdoped samples.  
The qualitative difference between the two scenarios is the behavior of the slope of the $B_{2g}$ Raman response -- the slope, measured in units of $R_0/\Delta$, varies with  doping via the varying gap anisotropy,
but is doping independent if the effect is due to final state interactions.
 
To extract the doping dependence of the slope from the data, 
one needs to normalize the theoretical results in the same way as in experiment.   Le Tacon \etal~\cite{gabi}  normalized the measured Raman response
 for all dopings by imposing an empirical `sum rule'
\begin{equation}
\int d\Omega \Omega~Im R_{B_{2g}} (\Omega) = C p
\label{n_1}
\end{equation}
where $p$ is the doping and $C$ is a constant.  They argued
 that this sum rule can be approximately derived for a 
weakly doped Mott insulator.   With this normalization, 
 the slope of $Im R_{B_{2g}}$ is essentially doping independent,
 although the peak positions in both the $B_{1g}$ and $B_{2g}$ channels vary substantially. 

The theoretical  Raman intensity contains a prefactor $R_0$ (see Eq.~(\ref{2})).
It is proportional to the fermionic density of states, $N$, and 
 the square of the quasiparticle residue, $Z$. Both $N$ and $Z$ 
 are weakly dependent on doping for overdoped samples, but
 become doping dependent in the strong coupling regime.
The normalization used by   Le Tacon \etal~implies that 
\begin{equation}
R_0 \Delta^2_0 \int dz z Im {\tilde R}_{B_{2g}} (z) = C p
\label{n_2}
\end{equation}
where $z = \Omega/\Delta$, and $Im {\tilde R} (z)$ is the function plotted in the figures.  

The slope of 
 $Im R_{B_{2g}} (z)$ scales as $R_0/\Delta$ if  the gap has a
 simple $d-$wave form, even if
final state interactions are relevant (the final state interaction does not affect the slope).  The doping independence of the slope then
 implies that $R_0 \sim \Delta$. Substituting this into Eq.~(\ref{n_2}), we 
find that this is consistent with the data if 
\begin{equation}
I (\Gamma_{B_{2g}}) = \int dz z Im {\tilde R}_{B_{2g}} (z) \propto \frac{p}{\Delta^3_0}
\label{n_3}
\end{equation}
At the doping dependence of $\Delta$ is known, Eq.~(\ref{n_3}) 
is a parameter-free condition that can confirm or disprove our theory. 
We computed the integral of $Im {\tilde R}_{B_{2g}}$ from Fig.~\ref{fig4_1}
 and found $I(0) \approx 0.94$, $I(-1) \approx 0.30$, and 
$I(-1.5) \approx 0.20$.
Extracting the peak position in units of $\Delta$ from  Fig.~\ref{fig4_1}
 and $\Delta$ from Fig. 2 of Ref.~\onlinecite{gabi}, we find that 
$\Gamma_{B_{2g}} =0$ roughly describes the optimal doped $T_c = 95K$ sample ($p=0.15$), 
$\Gamma_{B_{2g}} =-1$ the underdoped $T_c = 86K$ sample ($p=0.12$)
 for which $\Delta$ is about $1.4$ times larger than at optimal doping, and 
 $\Gamma_{B_{2g}} = -1.5$ the underdoped $T_c = 63K$ sample ($p=0.1$)
 for which $\Delta$ is about $1.5$ times larger than at optimal doping.
Compared to optimal doping, $p/\Delta^3$ in the two underdoped cases is reduced by 
$0.29$ and $0.20$, respectively. These numbers are  in good agreement with 
 our $I(-1)/I(0) =0.32 $ and $I(-1.5)/I(0) =0.21$. 
This good agreement is a strong argument for 
 an explanation  based on final state interactions rather than for a strong 
 doping-dependent  anisotropy of the gap.

It is instructive to further compare
our explanation of the data with the one
 presented by Le Tacon \etal~\cite{gabi}.  They included both a $\cos(6\phi)$ contribution to the gap $\Delta (\phi)$ (as in Eq.~(\ref{10})) 
 and a Fermi liquid renormalization, $\Lambda$, of the Raman vertex. 
They argued that the vertex renormalization factor $\Lambda$ 
is angle dependent,
$\Lambda = \Lambda_0(1 - C \cos^2 {2\phi})$, and that $C$  increases with underdoping.  For nodal fermions $(\phi = \pi/4)$, the vertex equals $\Lambda_0$ and is doping independent, but for antinodal fermions, its values goes down with underdoping.   Le Tacon \etal~argued that they needed the $C$ term to
explain the loss of intensity of the $B_{1g}$ peak, and  also to obtain a maximum in the $B_{2g}$ response, which in the absence of the $C$ term is flat.

In our theory, the intensity of the  the $B_{1g}$ peak
shifts down because of the final state interaction, which is equivalent to
the renormalization of the Raman vertex  The suppression in our case is 
the consequence of the fact that the intensity of the peak   scales as $\sqrt{-1/\Gamma}$. In this, we agree with  Le Tacon \etal. At the same time,  the $C$ term in the analysis by Le Tacon \etal~reduces the $B_{1g}$ vertex for all frequencies, while in our theory, the $B_{1g}$ Raman vertex at the smallest frequencies is enhanced by final state interactions if $\Gamma_{B_{1g}} <0$ 
(in Fig.~\ref{fig4}, the $B_{1g}$ intensity at small $\Omega$ is larger than the one without final state interactions). 

In another distinction from  Le Tacon \etal, the $B_{1g}$ peak in our theory  
 shifts down compared to $2\Delta$. However,  the energy of the peak 
 still increases with underdoping because $\Delta$ increases~\cite{girsh,cbm,cdk}.

For $B_{2g}$ scattering, the effect of the $C$ term in their consideration
 is  consistent with our analysis. Namely, their vertex becomes progessively smaller with the deviation of a  typical angle $\phi$ from $\pi/4$. This obviously happens as the frequency increases. In our case, the ratio of the
 renormalized vertex to the bare one 
 also progressively decreases as the frequency increases, if $\Gamma_{B_{2g}} <0$. From this perspective, the angular dependence of their $C$ term 
also  mimics
  the effect of our final state interaction in the $B_{2g}$ channel 
for a negative sign of $\Gamma_{B_{2g}}$.
 However, in distinction to Le Tacon \etal, we argue that this effect alone explains the data, i.e., there is no need to invoke a doping dependent change of the anisotropy of the $d-$wave gap.
  
To be more quantitative,  LaTacon \etal~used the experimental $\Delta$ 
and three other  doping dependent 
parameters: $Z\Lambda_0$, $C$, and the gap anisotropy parameter, $a$.  Two out of three of these parameters are fixed by (i) the `sum rule'
 requirement, and (ii) the experimental fact that the slope of the normalized
 Raman $B_{2g}$ response is doping independent. The one free 
 doping dependent parameter is chosen to fit the position of the $B_{2g}$ peak.
     
In our theory, we assume a simple $d-$wave gap and calculated 
 the vertex renormalization based on the RPA~\cite{RPA}. We therefore 
 have only two parameters: 
 $R_0$, and $\Gamma_{B_{2g}}$ to fit the same three sets of data.
  As we described above, we found that the agreement with the data 
  is nearly perfect in the sense that once we fix 
 $R_0$ and $\Gamma_{B_{2g}}$ to
match the slope and  the peak position in the 
$B_{2g}$ channel, we find that the experimental 
normalization condition is satisfied, despite the strong doping 
variation of $\Delta$.

We also note that the dependence on the fermionic damping in Figs.~\ref{fig8} and \ref{fig7} is consistent with the data if we assume, like in earlier work~\cite{mike_mod,elihu}, that $\eta$ increases strongly above $T_c$. Namely, the 
 maximum in the $B_{1g}$ intensity measured by Le Tacon \etal~not only shifts, but also  broadens with underdoping, and at small $\Omega$ 
the intensity above $T_c$ overshoots the intensity in the superconducting state.  For $B_{2g}$ scattering, the intensity in the superconducting state overshoots the intensity in the normal state in a broad range of frequencies, particularly
 for the most underdoped $T_c =63K$ sample (see Fig.~1 of Ref.~\onlinecite{gabi}).
This behavior is in  agreement with Fig.~\ref{fig7}.

An issue left  in our analysis and in the analysis by LeTacon \etal~is the justification of the sign of the effective interaction in the $B_{2g}$ channel 
(this is the sign of $\Gamma_{B_{2g}}$ in our analysis, and the sign of the $C$ term in the analysis of  LeTacon \etal).  To fit the data, we need 
$\Gamma_{B_{2g}} <0$.  As we said in Sec.~II, the full $\Gamma_{B_{2g}}$ has charge and spin components: $\Gamma_{B_{2g}} =  
\Gamma^c_{B_{2g}} +3 \Gamma^s_{B_{2g}}$.  The spin component $ \Gamma^s_{B_{2g}}$ has opposite sign compared to $ \Gamma^s_{B_{1g}}$ as the interaction 
 is peaked at or near the antiferromagnetic momentum $Q =(\pi,\pi)$.
 That is, 
 $\Gamma^s_{i} \propto \gamma_i (k) \gamma_i (p) \chi_s (k-p)$, where $k$ and $p$ are on the Fermi surface,  and the sign difference between the $B_{2g}$ and $B_{1g}$ channels is due to the fact that 
 $\gamma_{B_{1g}} (k +Q) = -\ \gamma_{B_{1g}} (k)$, while $\gamma_{B_{2g}} (k+Q) = \gamma_{B_{2g}} (k)$.
If $d-$wave pairing is magnetically mediated, $ \Gamma^s_{B_{1g}} <0$, hence
$ \Gamma^s_{B_{2g}}>0$ (from a pairing perspective, the latter is repulsive).
 However,  the dominant contribution to 
$\Gamma_s$ comes from regions near the hot spots, which in the cuprates are rather close to $(\pi,0)$, for which $\gamma_{B_{2g}}=0$.
 As a consequence, the magnitude of $ \Gamma^s_{B_{2g}}$ is substantially smaller than that of $ |\Gamma^s_{B_{1g}}|$. In this situation, the 
charge component of the $B_{2g}$ Raman vertex may well exceed the spin component. The charge and spin components of $\Gamma_i$ contribute with a different sign to the pairing channel (because of the spin factor $\sigma^y$ in the pairing vertex). Accordingly,  the charge-dominated repulsive pairing interaction in the 
$B_{2g}$ channel corresponds to  $\Gamma^c_{B_{2g}} <0$, and, hence, to 
 a negative $\Gamma_{B_{2g}}$, which we need.  In this context,  the 
increase in $|\Gamma_{B_{2g}}|$ with decreasing $p$ well may be
 a consequence of the  reduced screening when approaching the Mott transition.

\section{Conclusions}

In summary, we argued in this paper that the data on $B_{1g}$ and $B_{2g}$ 
 Raman scattering by Le Tacon \etal~\cite{gabi} can be explained as an effect of final state interactions, without the need to invoke a doping-dependent change in the anisotropy of the d-wave pairing gap.
 This work demonstrates that care must be taken when comparing energy gaps derived from
 two particle spectroscopies like Raman from those obtained directly from `single particle' probes
 such as photoemission.   

\acknowledgments

AVC acknowledges support from NSF-DMR 0604406, from the
 visitor program of the University of Chicago and Argonne National Laboratory, 
 and is thankful to TU-Braunschweig for
their hospitality during the initial stages of this work. 
 MRN was supported by the U.~S.~Dept.~of Energy, Office of Science, under Contract No.~DE-AC02-06CH11357. We would like to 
thank I. Eremin, M. Randeria, and J.C. Campuzano for useful discussions.


\begin{thebibliography}{99}

\bibitem{AP} M. R. Norman, D. Pines and C. Kallin, Adv. Phys. {\bf 54}, 715 (2005).

\bibitem{mike_mod} M. R. Norman, A. Kanigel, M. Randeria, U. Chatterjee and J. C. Campuzano, Phys. Rev. B {\bf 76}, 174501 (2007).

\bibitem{elihu} A. V. Chubukov, M. R. Norman, A. J. Millis and E. Abrahams, Phys. Rev. B {\bf 76}, 180501(R) (2007).

\bibitem{amit1} A. Kanigel, M. R. Norman, M. Randeria, U. Chatterjee, S. Souma, A. Kaminski, H. M. Fretwell, S. Rosenkranz, M. Shi, T. Sato, T. Takahashi, Z. Z. Li, H. Raffy, K. Kadowaki, D. Hinks, L. Ozyuzer and J. C. Campuzano, Nature Phys. {\bf 2}, 447 (2006).

\bibitem{amit2} A. Kanigel, U. Chatterjee, M. Randeria, M. R. Norman, S. Souma, M. Shi, Z. Z. Li, H. Raffy and J. C. Campuzano, Phys. Rev. Lett. {\bf 99}, 157001 (2007).

\bibitem{davis} T. Hanaguri, Y. Kohsaka, J. C. Davis, C. Lupien, I. Yamada, M. Azuma, M. Takano, K. Ohishi, M. Ono and H. Takagi, Nature Phys. {\bf 3}, 865 (2007); J. C. Davis, private communication.

\bibitem{mesot} J. Mesot, M. R. Norman, H. Ding, M. Randeria, J. C. Campuzano, A. Paramekanti, H. M. Fretwell, A. Kaminski, T. Takeuchi, T. Yokoya, T. Sato, T. Takahashi, T. Mochiku and K. Kadowaki, Phys. Rev. Lett. {\bf 83}, 840 (1999).

\bibitem{john} J. F. Zasadzinski, L. Ozyuzer, N. Miyakawa, K. E. Gray, D. G. Hinks and C. Kendziora, Phys. Rev. Lett. {\bf 87}, 067005 (2001).

\bibitem{gabi} M. Le Tacon, A. Sacuto, A. Georges, G. Kotliar, Y. Gallais, D. Colson and A. Forget, Nature Physics {\bf 2}, 537 (2006).

\bibitem{girsh} G. Blumberg, M. Kang, M. V. Klein, K. Kadowaki and C. Kendziora, Science {\bf 278}, 1427 (1997).

\bibitem{RMP} T. P. Devereaux and R. Hackl, Rev. Mod. Phys. {\bf 79}, 175 (2007).

\bibitem{devereaux} T. P. Devereaux and D. Einzel, Phys. Rev. B {\bf 51}, 16336 (1995).

\bibitem{klein} M. V. Klein and S. B. Dierker, Phys. Rev. B {\bf 29}, 4976 (1984).

\bibitem{foot1}
The figures are obtained without the constant term.  However, we have found 
that the inclusion of this term does not change the results in any substantial way.  

\bibitem{cbm} A. V. Chubukov, D. K. Morr and G. Blumberg, Solid State Comm. 
{\bf 112}, 183 (1999).

\bibitem{cdk} A. Chubukov, T. P. Devereaux and M. V. Klein, 
Phys. Rev. B {\bf 73}, 094512 (2006).

\bibitem{dahm} T. Dahm, D. Manske, and L. Tewordt, Phys. Rev. B {\bf 59}, 14740 (1999). 

\bibitem{lee} P. A. Lee, Phys. Rev. Lett. {\bf 71}, 1887 (1993).

\bibitem{RPA} It should be noted that within the RPA, factors such as $Z$ and $\Lambda$ do not
explicitly appear.  That is, one uses unrenormalized Greens functions, and the effect of
interactions is solely accounted for by $\Gamma$.

\end{thebibliography}
\end{document}